\documentclass[aps,prb,superscriptaddress,a4paper,11pt]{revtex4-1}

\usepackage[english]{babel}
\usepackage{graphicx}
\usepackage{amsmath}
\usepackage{amssymb}
\usepackage{color,soul}

\newcommand{\buenosaires}{Universidad de Buenos Aires, 
Facultad de Ciencias Exactas y Naturales, 
Departamento de F\'isica, 
C1428EHA Buenos Aires, Argentina}
\newcommand{\CONICET}{CONICET - Universidad de Buenos Aires, 
Instituto de F\'isica de Buenos Aires (IFIBA), 
C1428EHA Buenos Aires, Argentina}

\begin{document}
\title{Case study of the validity of truncation schemes of kinetic 
equations of motion: few magnetic impurities in a semiconductor 
quantum ring}

\author{J.\ M.\ Lia}
\affiliation{\buenosaires}
\affiliation{\CONICET}

\author{P.\ I.\ Tamborenea}
\affiliation{\buenosaires}
\affiliation{\CONICET}

\date{\today}

\begin{abstract}
We carry out a study on the validity and limitations of truncation 
schemes customarily employed to treat the quantum kinetic equations 
of motion of complex interacting systems.
Our system of choice is a semiconductor quantum ring with one electron
interacting with few magnetic impurities via a Kondo-like Hamiltonian.
This system is an interesting prototype which displays the necessary
complexity when suitably scaled (large number of magnetic impurities) 
but can also be solved exactly when few impurities are present.
The complexity in this system comes from the indirect 
electron-mediated impurity-impurity interaction and is reflected in 
the Heisenberg equations of motion, which form an infinite hierarchy.
For the cases of two and three magnetic impurities, we solve for 
the quantum dynamics of our system both exactly and following a 
truncation scheme developed for diluted magnetic semiconductors 
in the bulk.
We find an excellent agreement between the two approaches when 
physical observables like the impurities' spin angular momentum 
are computed for times that well exceed the time window of validity 
of perturbation theory.
On the other hand, we find that within time ranges of physical 
interest, the truncation scheme introduces negative populations 
which represents a serious methodological drawback.

\end{abstract}

\maketitle

\section{Introduction}

Many-body interacting systems such as diluted magnetic semiconductors 
(DMS) pose interesting theoretical challenges.
Their quantum dynamics is completely described by the Heisenberg 
equations of motion for the density matrix.
These equations are usually coupled to one another and an analytical 
solution is in general not always possible.
In different areas of physics, there exists a long tradition of 
approaching the study of this kind of systems of equations of motion 
by ordering them into a hierarchy of increasing correlations between 
the particles or the fields
involved.\cite{thu-axt,aar-bon-wet,ley-foe-wie,xu-yan}
The work of Kubo\cite{kub} on the expansion of cumulant functions for
stochastic variables has proven useful in carrying out this reordering, 
and the ideas expounded in his paper have been applied to problems of
condensed matter physics
such as that of optical excitation in
semiconductors\cite{axt-sta,ros-kuh,khi-gib-jan,lin-hu-bin} 
and, more recently,
the theoretical treatment of DMS.\cite{thu-axt,ung-cyg-axt,ung-cyg-axt-2}
Typically, only approximate solutions to the system of equations 
can be obtained. 
Once the relevant hierarchy has been established, it is truncated
following a particular scheme that discards high order correlations 
and leads to another set of equations that is at least numerically 
tractable.\cite{ros-kuh}

In the context of DMS, the study of nanostructures is attracting
growing interest.\cite{kac,bli-kac,mor,cha-li-xia,wu-jia-wen,ma,
  kra-vla-sca,ung-cyg-axt-4,vie-kos-sin,die,die-ohn}
Among these structures are narrow quantum rings (QR) 
\cite{fru-ric} with few 
magnetic impurities which, due to their simplicity 
and experimental feasibility,\cite{vie-kos-sin,yak-mer} 
are particularly well suited for exploring the strengths and limitations
of truncation schemes.
When the number of impurities is small, the ultrafast quantum dynamics 
of these systems can be computed exactly without resorting to the 
Heisenberg equations.
Such exact solutions are useful since they can be used as
benchmarks to which the approximate solutions coming from
truncation schemes can be compared.
Thus, here we pose the quantum dynamics problem of a DMS QR modelled
with the Kondo interaction\cite{kon} between the electron spin and 
the magnetic impurities.
Our purpose is twofold: on the one hand, we wish to further our 
studies of angular momentum dynamics and control in 
nanostructures\cite{lia-tam,lia-tam-cyg}.
On the other hand, and more to the point of this article, we report
in a quantitative way the encountered methodological difficulties,
in order to contribute to the development and improvement of 
theoretical techniques based on hierarchies of equations of motion.

The paper is organized as follows.
In Sec.\, \ref{sec:quantum_ring_system} we lay out the steps and 
assumptions leading to the one-dimensional model for the DMS QR to 
which we devote this study.
In Sec.\,\ref{sec:trunc_scheme} we describe at length the truncation 
scheme that we apply to the Heisenberg equations for the many-body 
density matrices.
In Sec.\,\ref{sec:num_results} we integrate numerically the 
truncated Heisenberg equations and, when possible, compare the 
results with their exact counterparts, which are computed by solving 
the time-dependent Schr\"odinger equation.
Finally, in Sec.\,\ref{sec:conclusions} we offer some concluding 
remarks.

%%%%%%%%%%%%%%%%%%%%%%%%%%%%%%%%%%%%%%%%%%%%%%%%%%%%%%%%%%%%%%%%%%%%%%%
%%%%%%%%%%%%%%%%%%%%%%%%%%%%%%%%%%%%%%%%%%%%%%%%%%%%%%%%%%%%%%%%%%%%%%%

\section{Quantum ring system}
\label{sec:quantum_ring_system}

We consider a narrow semiconductor quantum ring doped with a single 
electron and a few Mn impurities.
In the envelope-function approximation, the Hamiltonian of the bare QR,
including the confining potential $U(\mathbf{r})$, reads
\begin{equation}
  H_0 = -\frac{\hbar^2}{2m^\ast}\nabla^2 + U(\mathbf{r})
  \label{eq:H0_3d}
\end{equation}
where $m^\ast$ is the conduction-band effective mass.
Between the electron and the \textit{d}-shell spin of the impurities 
we assume the typical \textit{sd} exchange interaction described by
the Kondo-like Hamiltonian\cite{thu-axt,qu-haw}
\begin{equation}
  H_{sd} = J\sum_{I=1}^{N} \mathbf{S}_I\cdot\mathbf{s}\,
  \delta(\mathbf{r} - \mathbf{R}_I)
  \label{eq:Hsd_3d}
\end{equation}
where $N$ is the number of Mn impurities, $J$ the bulk \textit{sd} 
exchange constant, \(\mathbf{s}\) the spin of the electron, 
and \(\mathbf{S}_I\) and \(\mathbf{R}_I\) the spin and position 
of the $I$-th impurity, respectively.
Note that $H_{sd}$ conserves the total spin angular momentum (SAM),
$\mathbf{s} + \sum_{I=1}^{N} \mathbf{S}_I$.\cite{thu-axt}

Here we adopt a quasi-one-dimensional model for the QR in which 
the radial and vertical components of the wave function are taken 
as the respective ground 
states\cite{mei-mor-klap,lor-joh-kot,lin-lin-ling}
and do not participate in the dynamics.
The resulting $\varphi$-dependent Hamiltonian reads
\begin{equation}
  H = \frac{E_0}{\hbar^2}L_z^2 +
  \frac{J}{V}\sum_{i=1}^{N}
  \mathbf{s}\cdot\mathbf{S}_I\, \delta(\varphi-\varphi_I);
  \label{eq:H}
\end{equation}
where $L_z=-i\hbar\partial_{\varphi}$ is the $z$-component operator 
of the electron's orbital angular momentum (OAM),
$E_0=\hbar^2/2m^{\ast}a^2$, with
$a$ being the radius of the ring, and $V$ is the volume of the QR.
The location of the impurities is specified by the angular 
variables $\varphi_I$.

The time evolution driven by the many-body Hamiltonian of 
Eq.\ \eqref{eq:H} can be obtained numerically by solving 
the Schr\"odinger equation if $N$ is sufficiently  small.
For large $N$ (say $N>4$) this is no longer practical or even
possible.
In such cases one resorts to the equations of motion of the 
density matrices, which form a coupled and infinite hierarchy.
Here the pitfall is that only by truncating this hierarchy 
a numerically tractable closed set of equations can be obtained.
The question is: How to carry out the truncation while preserving 
both the basic mathematical properties of the density matrices and 
the fundamental physical features of the model?

In this work we follow a well-established procedure to treat the 
hierarchy of quantum density matrices \cite{thu-axt,kub} and 
apply it to the QR with one electron and a few magnetic impurities.
For bulk DMS, this method yields good approximate solutions on 
short time scales, which preserve the fundamental symmetries and 
their associated conserved quantities.
However, on longer time scales (e.g., beyond the regime of 
perturbation theory), its performance has not been sufficiently 
explored.
Here we test this methodology in a rather small version of a DMS 
system, taking advantage of the fact that we can compare its results 
for long times with exact solutions of the Hamiltonian evolution.

%%%%%%%%%%%%%%%%%%%%%%%%%%%%%%%%%%%%%%%%%%%%%%%%%%%%%%%%%%%%%%%%%%%%%%%
%%%%%%%%%%%%%%%%%%%%%%%%%%%%%%%%%%%%%%%%%%%%%%%%%%%%%%%%%%%%%%%%%%%%%%%

\section{Truncation Scheme}
\label{sec:trunc_scheme}

In terms of many-body operators the Hamiltonian in Eq. \eqref{eq:H} 
reads
\begin{equation}
  H = E_0\sum_{m\sigma} m^2\,c_{m\sigma}^{\dag}c_{m\sigma} + \frac{J}{V}
  \sum_{\substack{Inn'\\m\sigma m'\sigma'}}
  \mathbf{s}_{\sigma\sigma'} \cdot \mathbf{S}_{nn'}\,\rho_{mm'}^{I} \,
  c_{m\sigma}^{\dag}c_{m'\sigma'}P_{nn'}^{I}.
  \label{eq:H_2q}
\end{equation}
In this expression $\mathbf{s}_{\sigma\sigma'}$ are the matrix 
elements of the
electron's spin operator in the basis of eigenstates of 
$s_z$ ($\sigma=\pm1/2$),
and $\rho_{mm'}^{I}\doteq e^{i(m-m')\varphi_I}$ 
are the matrix elements of the
delta function at $\varphi_I$ in the basis of eigenvalues of the 
$L_z$ operator ($m\in\mathbb{Z}$).
The operators $P_{nn'}^{I}\doteq|I,n\rangle\langle I,n'|$ are 
defined through the equations
\begin{equation}
  \mathbf{S}_I = \sum_{nn'} \langle I, n|\mathbf{S}_{I}|I, 
  n'\rangle P_{nn'}^{I},
  \label{eq:P_def}
\end{equation}
where $|I, n\rangle$, $n\in\{-5/2,-3/2,\ldots,3/2,5/2\}$, are 
the eigenstates of the spin 5/2 operator $S_{z}^{I}$.
The $P^{I}$ operators are therefore interpreted as density matrices.
Notice that $[P^{I}, P^{I'}]=0$ for $I\neq I'$, since they act 
on different impurities, but
$[P_{n_1n_2}^{I},P_{n_3n_4}^{I}] = P^I_{n_1n_4}\delta_{n_2n_3} -
P^I_{n_3n_2}\delta_{n_1n_4}$.

Derived from the Hamiltonian in Eq. \eqref{eq:H_2q}, the Heisenberg
equations of motion for the expectation values
$\langle c_{m_1\sigma_1}^{\dag} c_{m_2\sigma_2}\rangle$ and
$\langle P_{n_1n_2}^{I}\rangle$ read
\begin{align}
 \begin{split}
  i\hbar\frac{\partial}{\partial t}\langle P^{I}_{n_1n_2}\rangle 
  &= \sum_{\substack{nm\sigma\\m'\sigma'}}
  \rho_{mm'}^{I} \mathbf{s}_{\sigma\sigma'}\cdot\left(\mathbf{S}_{n_2n} \langle
    c^{\dag}_{m\sigma}c_{m'\sigma'}P_{n_1n}^{I}\rangle - \mathbf{S}_{nn_1} \langle
  c^{\dag}_{m\sigma}c_{m'\sigma'}P_{nn_2}^{I}\rangle \right)
 \end{split}
 \label{eq:cc_P_P}
 \\
 \begin{split}
   i\hbar\frac{\partial}{\partial t}\langle
   c_{m_1\sigma_1}^{\dag}c_{m_2\sigma_2}\rangle &= E_0(m_2^2 - m_1^2)
   \langle c_{m_1\sigma_1}^{\dag}c_{m_2\sigma_2}\rangle \\
   &\quad +
   \sum_{\substack{Inn'\\m\sigma}} \mathbf{S}_{nn'} \cdot \left(
   \rho_{mm_1}^{I} \mathbf{s}_{\sigma\sigma_1} \langle
   c^{\dag}_{m\sigma}c_{m_2\sigma_2}P_{nn'}^{I}\rangle
   -\rho_{m_2m}^{I}\mathbf{s}_{\sigma_2\sigma} \langle
   c^{\dag}_{m_1\sigma_1}c_{m\sigma}P_{nn'}^{I}\rangle \right).
 \end{split}
 \label{eq:cc_P_cc}
\end{align}
The dynamics introduced by the \textit{sd} interaction in the 
two-point density matrices for the electron and each impurity 
in the system therefore depend solely on the three-point 
matrices $\langle c^{\dag}cP^{I}\rangle$.
Instead of truncating the hierarchy at this level, we take one 
step further and add to Eqs. \eqref{eq:cc_P_P} and \eqref{eq:cc_P_cc} 
the equations of motion for
$\langle c^{\dag}cP^{I}\rangle$, which we express as
\begin{equation}
  \begin{split}
  i\hbar\frac{\partial}{\partial t} \langle
  c^{\dag}_{m_1\sigma_1} c_{m_2\sigma_2} P_{n_1n_2}^{I}\rangle = &E_0(m_2^2 -
  m_1^2)\langle c_{m_1\sigma_1}^{\dag} c_{m_2\sigma_2} P_{n_1n_2}^{I}\rangle \\
    &+ \frac{J}{V} \sum_{nm\sigma}
      \left(\mathbf{S}_{n_2n}\cdot\mathbf{s}_{\sigma_2\sigma} 
    \rho_{mm_2}^{I} \langle c^{\dag}_{m_1\sigma_1} c_{m\sigma} 
    P_{n_1n}^{I} \rangle\right. \\
    &- \left.
    \mathbf{S}_{nn_1}\cdot\mathbf{s}_{\sigma\sigma_1} 
    \rho_{mm_2}^{I} \langle c^{\dag}_{m\sigma} c_{m_2\sigma_2}
    P_{nn_2}^{I} \rangle\right) + Q;
  \end{split}
  \label{eq:ccP}
\end{equation}
where the term $Q$ (actually $Q^I_{m_1\sigma_1m_2\sigma_2n_1n_2}$) 
collects all contributions from four-point density matrices, 
including those of the indirect
interaction between the impurities, and is defined as follows
\begin{equation}
  Q = \frac{J}{V}\sum_{\substack{I\neq I'\\nn'm\sigma}} 
  \mathbf{S}_{nn'}\cdot\left(\mathbf{s}_{\sigma_2\sigma}
    \rho_{mm_2}^{I'} \langle c^{\dag}_{m_1\sigma_1} c_{m\sigma} 
    P_{n_1n_2}^{I}P_{nn'}^{I'} \rangle - \mathbf{s}_{\sigma\sigma_1}
    \rho_{mm_2}^{I'} \langle c^{\dag}_{m\sigma} c_{m_2\sigma_2}
    P_{n_1n_2}^{I} P_{nn'}^{I'} \rangle\right).
  \label{eq:Q}
\end{equation}
When only one impurity is present $Q$ vanishes identically and 
the hierarchy does not develop further.
In this case the set comprising Eqs.\ \eqref{eq:cc_P_P}, 
\eqref{eq:cc_P_cc} and \eqref{eq:ccP} is closed and the 
eigendecomposition of the full Hamiltonian can be worked out 
exactly without resorting to numerical methods.\cite{she-cha-1}

Let us now truncate the hierarchy so as to obtain a set of 
equations that is closed at the three-point level.
In order to do that we first apply the expansion described in
Ref.\ [\onlinecite{kub}] to each of the four-point density 
matrices appearing in $Q$, and rewrite them as
\begin{equation}
  \begin{split}
    \langle c^{\dag}_{m_1\sigma_1} c_{m_2\sigma_2} P^{I} P^{I'} \rangle =
    &\langle P^{I}\rangle \langle c^{\dag}_{m_1\sigma_1} c_{m_2\sigma_2}
      P^{I'}\rangle +
    \langle P^{I'}\rangle \langle c^{\dag}_{m_1\sigma_1} c_{m_2\sigma_2}
    P^{I}\rangle
    +\langle c^{\dag}_{m_1\sigma_1} c_{m_2\sigma_2}\rangle
      \delta\langle P^{I} P^{I'}\rangle
    \\
    &-\langle c^{\dag}_{m_1\sigma_1} c_{m_2\sigma_2}\rangle \langle P^{I}\rangle
    \langle P^{I'}\rangle + \delta\langle c^{\dag}_{m_1\sigma_1} c_{m_2\sigma_2}
      P^{I} P^{I'}\rangle;
 \end{split}
  \label{eq:ccPP_expansion}
\end{equation}
where we omit the subindices of the operators $P^{I}$ and $P^{I'}$ 
for clarity.
In this expression the factor $\delta\langle P^{I}P^{I'}\rangle$ 
is defined as
$\delta\langle P^{I}P^{I'}\rangle \doteq \langle P^{I}P^{I'}
\rangle - \langle P^{I}\rangle \langle P^{I'}\rangle$, 
and the rightmost term contains, 
by definition, all contributions to the left-hand-side that are not
reducible to a factorized form similar to those of the other terms.
Notice that the expansion in Eq.\ \eqref{eq:ccPP_expansion} is 
exact as long as
we do not neglect any term;\cite{kub,thu-axt} and it is also 
symmetric with respect to the indices $I$ and $I'$ that label 
the impurities, since, by definition, $P^{I}$ and $P^{I'}$ 
commute when $I\neq I'$.
Furthemore, it follows from the definition of $P^{I}$ that it 
only makes sense to consider the case $I\neq I'$, because a 
four-point density matrix of the form
$\langle c^{\dag}_{m_1\sigma_1} c_{m_2\sigma_2} P^{I} P^{I} \rangle$ reduces to a three-point one containing only one operator $P^I$.
Rewriting $Q$ using the expansion in Eq.\ \eqref{eq:ccPP_expansion} 
makes explicit the contributions of the irreducible terms
$\delta\langle P^{I}P^{I'}\rangle$ and
$\delta\langle c^{\dag}_{\lambda_1} c_{\lambda_2} P^{I} 
P^{I'}\rangle$ to the dynamics of the system.
Finally, to truncate the hierarchy at the three-point level 
we only need to neglect the latter (see Appendix 
\ref{sec:appendix_PPp}).

We remark at this point that truncating the hierarchy at the 
two-point level yields a set of equations that can be computed 
directly from a mean-field Hamiltonian.
The resulting set of equations is obtained by substituting all 
three-point matrices $\langle c^{\dag}cP^{I}\rangle$ in 
Eqs.\ \eqref{eq:cc_P_P} and
\eqref{eq:cc_P_cc} for their mean-field factorizations
$\langle c^{\dag}_{\lambda_1} c_{\lambda_2}\rangle 
\langle P^I \rangle$.\cite{thu-axt}
In any case, it is worth emphasizing that, regardless of the 
level at which the hierarchy is truncated, the approximation 
is performed on the density matrices and not on the Hamiltonian 
itself.

In the following section we analyse how relevant these correlations 
are to the dynamics when $N$ is small and the system is initially 
in a pure state.
We purposely choose configurations that are numerically tractable 
in the Schr\"odinger picture in order to have a reliable reference 
solution to which the approximate ones can be compared.

%%%%%%%%%%%%%%%%%%%%%%%%%%%%%%%%%%%%%%%%%%%%%%%%%%%%%%%%%%%%%%%%%%%%%%%%%%%%%%%
%%%%%%%%%%%%%%%%%%%%%%%%%%%%%%%%%%%%%%%%%%%%%%%%%%%%%%%%%%%%%%%%%%%%%%%%%%%%%%%
%%%%%%%%%%%%%%%%%%%%%%%%%%%%%%%%%%%%%%%%%%%%%%%%%%%%%%%%%%%%%%%%%%%%%%%%%%%%%%%

\section{Numerical results}
\label{sec:num_results}

Let us consider a $\text{Zn}_{1-x}\text{Mn}_{x}\text{Te}$ QR in 
the highly-diluted limit 
$x\ll 1$ ($N/V\approx10^{-3}\,\text{nm}^{-3}$, where
$V\approx 777\text{ nm}^3$ is the volume of the ring and $N=2,3$ 
the number of impurities we consider in this study.)
To compute the ring's volume we assume an average height of
$1.5$~nm,\cite{lin-lin-ling} an inner radius of 
$a=14$~nm,\cite{lor-joh-kot} and
an effective and experimentally feasible width of 
approximately 8.4~nm.
The latter parameter is estimated using a well-known model 
that assumes a parabolic radial confining 
potential.\cite{cha-pie,lin-lin-ling,sha-sza-esm}
In the highly-diluted limit, the bulk \textit{sd} exchange 
constant for ZnTe is found to be $J=11\text{ meV nm}^3$ and 
largely independent of the number of impurities.\cite{fur}
This value in the bulk yields an effective coupling constant 
of $J/V\approx0.0142\text{ meV}$.
We also assume that in the highly-diluted limit the conduction-band
effective mass of the (Zn,Mn)Te does not differ considerably from 
that of pure ZnTe,
$m^{\ast}=0.2m_e$, where $m_e$ the bare electron mass.
For the radius considered this effective mass yields a 
conduction-band energy scale of $E_0\approx0.972$~meV, which is 
almost two orders of magnitude larger
than the effective \textit{sd} coupling.
Because $E_0\gg J/V$, the energy of the first excited radial 
state is expected to be far above that of the ground 
state $R_0$,\cite{lia-tam-cyg}
and the quasi-one-dimensional approximation is still valid, even though 
the effective ring width is of the order of its effective radius.

In the bulk the impurities in a highly-diluted DMS are expected 
to be quite separated from each other.
Even though the precise locations of the impurities cannot be 
predicted during fabrication, in order to reproduce this condition 
in the ring as accurately as possible we assume that they are 
maximally separated from one another.
In other words, we distribute them on the ring so that they form 
an $N$-sided regular polygon when $N>2$, and are diametrically 
opposite when $N=2$.

To carry out the numerical calculations we consider a sufficiently 
large basis of electronic states with a maximum energy of $25\,E_0$.
We assume in all cases that the electron initially occupies a 
state of low energy (of the order of $E_0$) with definite SAM 
and OAM.
Similarly, and for the sake of concreteness, we assume that each 
impurity is initially polarized on the $xz$ plane and aligned 
at angle $\beta$ from the ring's axis.
Such single-impurity states can be written as 
$d^{(5/2)}(\beta)|S_z;5/2\rangle$,
where $d^{(5/2)}(\beta)$ is the Wigner small $d$ matrix for 
spin $5/2$ and $|S_z;5/2\rangle$ the eigenstate of the $S_z$ 
operator of maximum projection.
Notice that an initial polarization on any other plane containing 
the ring's axis would describe the same dynamics if the electron 
is initially in an $s_z$ eigenstate, because the full Hamiltonian 
is a scalar operator with respect to rotations of the total SAM.
These initial conditions on the electron and the impurities states
can be met experimentally.
The former using twisted-light laser
beams\cite{qui-tam,qui-tam-ber,qui-tam-kuh,mik-sza-foe}, and 
the latter using suitable magnetic fields.
In other words, we assume that the initial state of the whole 
system (electron plus impurities) is a ket of the form
$|m\sigma\rangle|\text{Mn}_1\rangle\cdots|\text{Mn}_N\rangle$, where
$|m\sigma\rangle$ is the initial state of the electron, and
$|\text{Mn}_I\rangle$ is the initial state of the $I$-th impurity.
At the onset of the dynamics no entanglement therefore exists between 
the electron and the impurities or between the impurities themselves.
The two- and three-point density matrices 
$\langle P^IP^{I'}\rangle$ and
$\langle c_{m_1\sigma_1}^{\dag}c_{m_2\sigma_2}P_{n_1n_2}^I\rangle$ 
are therefore equal to their mean-field factorizations and their 
respective correlated parts are zero.
Finally, we integrate the equations on a time scale that is 
of the order of ultrafast interactions between photocarriers 
and impurities in DMS.\cite{kne-yak-bay,die-pey-gri,koe-mer-yak}

\begin{figure}[htbp]
  \centering
  \includegraphics[scale=0.7]{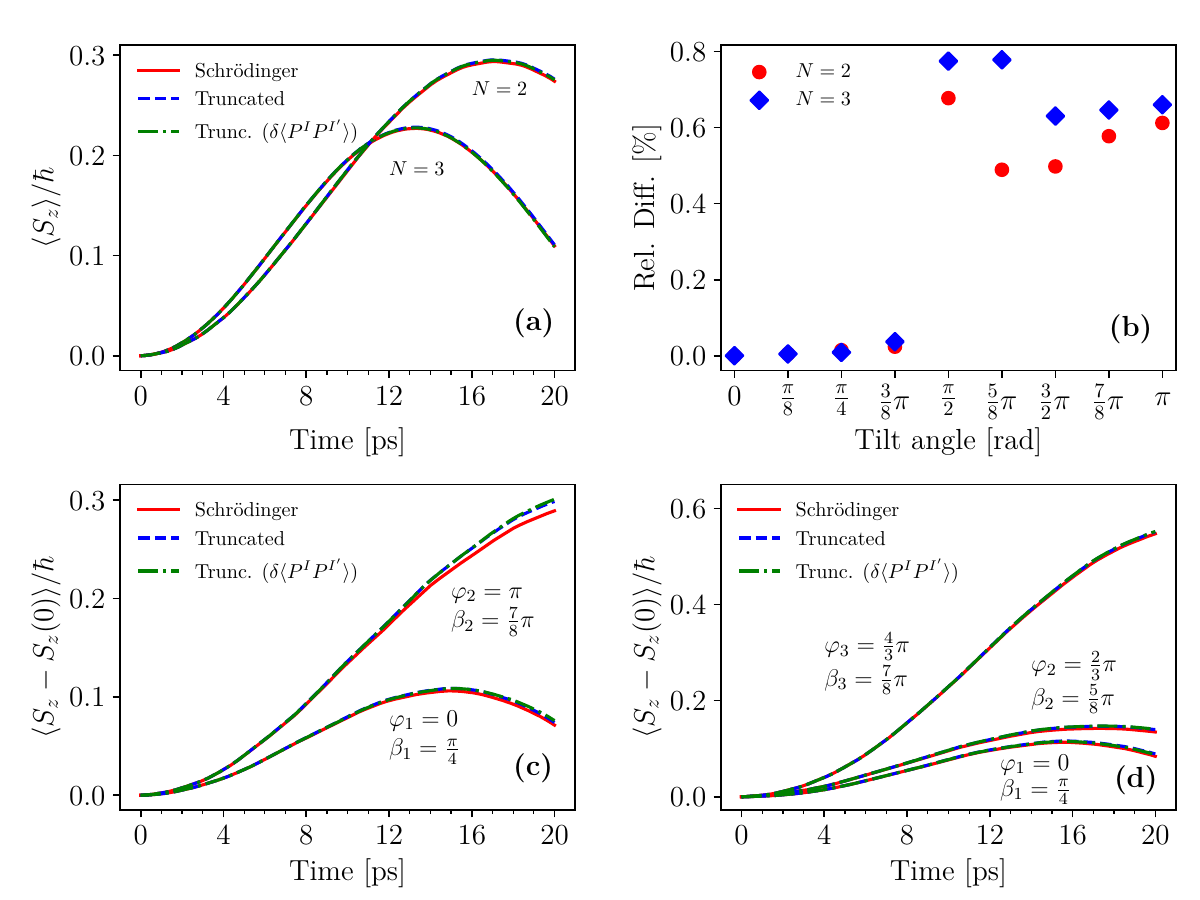}
  \caption{(a): Time evolution of the impurities' magnetization along 
  the ring's axis for an initial polarization of $5\hbar/2$ along 
  the $x$ axis. 
  The solid curve correspond to the exact (Schr\"odinger) solution, 
  while the dashed and dash-dotted lines to those obtained using 
  the truncation scheme including and leaving out the correlations 
  $\delta\langle P^IP^{I'}\rangle$. 
  (b): Maximum difference between the approximate and the referece 
  solutions, computed relative to the latter, for different tilt 
  angles and without the $\delta\langle P^IP^{I'}\rangle$. 
  (c-d): Same as (a) but assuming a different initial tilt angle 
  $\beta_I$ for each of the $N=2$ (c) or $N=3$ (d) impurities 
  located at $\varphi_I$. 
  For clarity, the magnetization is shown relative to its initial 
  value. 
  In all cases, the electron starts in the state $|1\uparrow\rangle$.}
  \label{fig:fig_1}
\end{figure}

Let us first assume that all the impurities' spins have been maximally 
polarized along the $x$ axis, that is $\beta=\pi/2$ and 
$\langle S_x^I\rangle=5\hbar/2$
and $\langle S_{y,z}^I\rangle=0$ for all $I$.
In Fig.\,\ref{fig:fig_1}a we pick one of the $N$ impurities in the system 
and display the time evolution of its magnetization along the ring's axis.
Which impurity we pick is immaterial, since all of them show the same spin
dynamics as a consequence of their highly symmetrical spatial distribution 
on the ring (see Appendix \ref{sec:appendix_polygon}).
The solid line corresponds to the reference (exact) solution obtained 
in the Schr\"odinger picture, while the dashed and dash-dotted lines 
show the dynamics of the same quantity as described by the truncation 
scheme with and without the direct impurity-impurity correlation 
$\delta\langle P^IP^{I'}\rangle$, respectively.
We see that in the time range considered, the approximate solutions 
including and leaving out this latter correlation are in excellent 
agreement with one another and each with the reference solution.
When the spins are aligned at different angles, the approximate solutions
differ from the reference in no more than $1\%$ when their separation reaches
the global maximum (Fig.\,\ref{fig:fig_1}b).
The same close correspondence is observed for a variety of randomly chosen
initial states, as well as for the case in which the impurities' spins are
aligned at different tilt angles.
A particular example of this case is presented in Fig.\,\ref{fig:fig_1}c for
$N=2$ and in Fig.\,\ref{fig:fig_1}d for $N=3$.
The addition of the equations for $\langle P^IP^{I'}\rangle$ to the 
original set is of no consequence at all, as expected in the 
highly-diluted limit.
We observe that in these situations, and when the average distance between the
manganese atoms is large enough, the impurity-impurity exchange terms may be
safely approximated by their mean-field contributions.

\begin{figure}[htbp]
  \centering
  \includegraphics[scale=0.7]{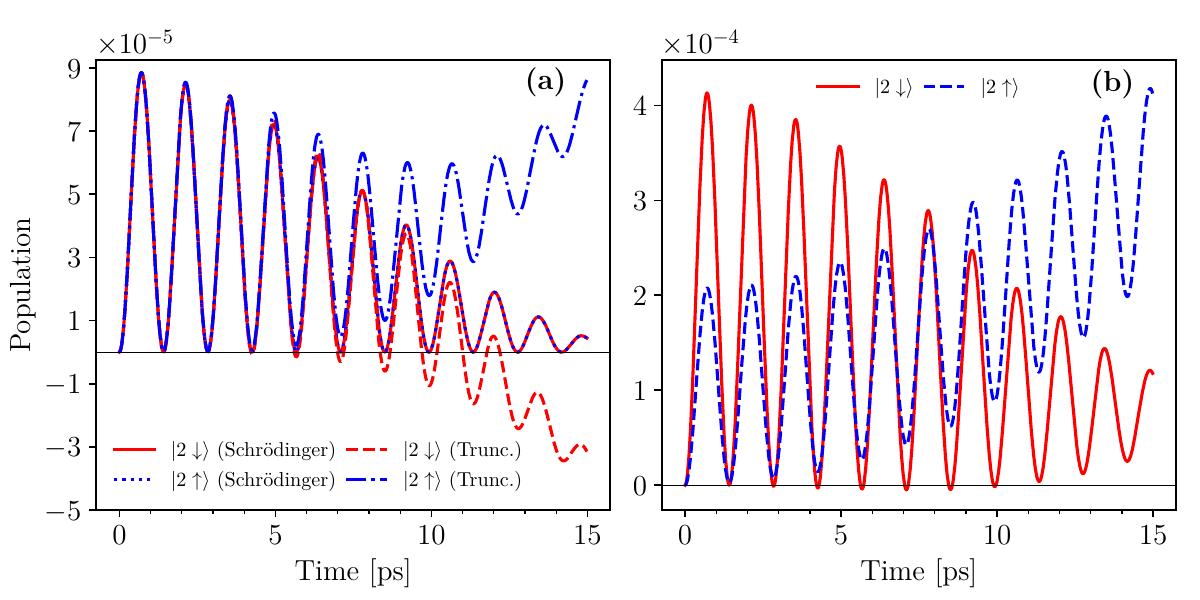}
  \caption{(a): Time evolution of the population of the electronic states
    $|2\uparrow\rangle$ and $|2\downarrow\rangle$ for the exact solution (solid
    lines) and the approximate one (dashed lines) not considering the
    contributions of the $\delta\langle P^IP^{I'}\rangle$. We assume that,
    initially, all impurities' spins are equally and maximally polarized along
    the $x$ axis ($\beta=\pi/2$).  (b): Population of the same electronic states
    but for the case in which the impurities' spins are initially oriented at
    random. In both figures we consider $N=3$ and the same initial electron
    state as in Fig.\,\ref{fig:fig_1}.}
  \label{fig:fig_2}
\end{figure}

The truncation scheme is, however, not so accurate in approximating the
transitions into and out of the available electron states.
For the impurities initially in the $S_x$ eigenstate of maximum projection
($\beta=\pi/2$) and an electron in the $|1\uparrow\rangle$ state, the
populations of the states $|2\uparrow\rangle$ and $|2\downarrow\rangle$ are
over- and underestimated throughout the time range considered
(Fig.\,\ref{fig:fig_2}a), respectively.
The discrepancy is worsened by the fact that the approximated populations
eventually take on negative values that, because of their magnitude, cannot be
ascribed to numerical error.
This behavior is observed as well for smaller integration time steps and
different initial states for the electron and the impurities' spins.
In particular, it is observed when the latter are oriented at random; that is,
when their state is initially described by the condition
$P^I_{n_1n_2}(t=0) = \frac{1}{6}\delta_{n_1n_2}$ for all $I$
(Fig.\,\ref{fig:fig_2}b).
In treating this case we make the additional assumption that correlations
$\delta\langle c_{m_1\sigma_1}^{\dag}c_{m_2\sigma_2}P^I_{n_1n_2}\rangle$ take
time to build up\cite{thu-axt} and are therefore initially zero (that is, the
electron's and the impurities' states are not initially entangled.)
However, regardless of the initial condition considered, the approximation always
respects the hermiticy of all two- and three-point density matrices in the set
of truncated equations.
Negative values for the populations therefore indicate that the positive
semi-definiteness of the electronic density matrix is not conserved during time
evolution.
\begin{figure}[htbp]
  \centering
  \includegraphics[scale=0.7]{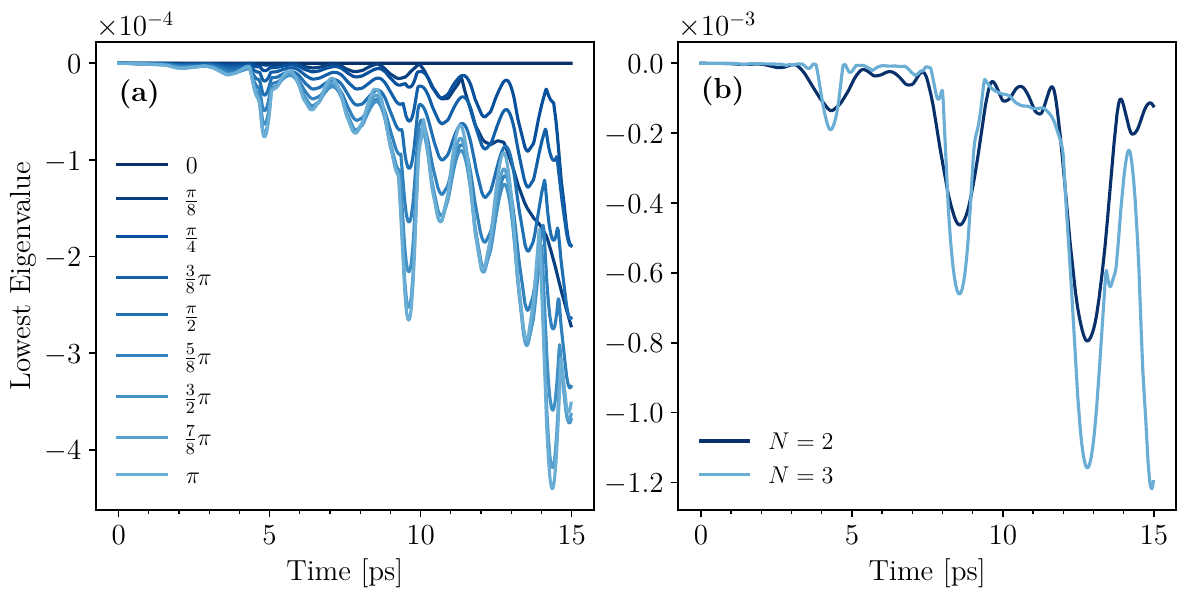}
  \caption{(a): Lowest eigenvalue of the electronic density matrix
    $\langle c_{m_1\sigma_1}^{\dag}c_{m_2\sigma_2}\rangle$ for the $N=3$ case
    and all impurities' spins initially on the $xz$ plane and oriented at the
    same tilt angle with respect to the ring's axis. (b): Same as (a) but for
    the randomly oriented ensemble. We assume the electron in the same initial
    state as in Fig.\,\ref{fig:fig_2}.}
  \label{fig:fig_3}
\end{figure}
This is revealed by the sign of its lowest eigenvalue, which is negative in all
but one of the cases presented in Figs.\,\ref{fig:fig_3}a-b.
In fact, the electronic density matrix becomes indefinite right at the first
integration step not only for the particular case $\beta=\pi/2$, but also for
other initial tilt angles (Fig.\,\ref{fig:fig_3}a), as well as for the case in
which the impurities' spins are oriented completely at random
(Fig.\,\ref{fig:fig_3}b).
%
%%%
Notice that the hermiticity of each truncated density matrix can be guaranteed
directly on the right-hand side of its equation of motion, since this property
depends at most on the density matrices themselves.
In contrast, their positive semi-definiteness requires a condition on their
second time derivative, and therefore depends strongly on $Q$.
The difference between both properties is most clearly reflected in the negative
populations shown in Figs.\,\ref{fig:fig_2}a-b.
Hermiticity of the whole density matrix requires the populations to be real (not
necessarily positive), but its positive semi-definiteness requires in addition
that they reach a local minimum whenever they become zero.
%
% The fact that the latter does not happen for some populations shows that their
% second time derivative at such points is negative.
%
From Eq. \eqref{eq:cc_P_cc}, it is not hard to see that this condition depends
on the first time derivative of the quantities
$\langle c_{m_1\sigma_1}^{\dag}c_{m_2\sigma_2}P^I_{n_1n_2}\rangle$ and therefore
directly on the truncated term $Q$.
This is also the case for the other density matrices in the truncated set.
Adding the direct impurity-impurity exchange term only introduces small
corrections to the values of the populations but does not help at all to solve
or reduce this problem.
The $\langle P^I\rangle$ matrices also lose their initial positive
semi-definiteness as their elements evolve in time.
It is only when the impurities' spins are maximally polarized along the axis of
the ring that this problem does not arise.
When this happens the spin part of the initial ket is the eigenstate of maximum
projection ($5N/2+1/2$) of the total SAM, which is a conserved quantity.
Neither the electron's nor the impurities' spins therefore change in time, since
there are no other available states for them to flip to while keeping the
maximum projection constant.
Whether correlations of the electron-mediated impurity-impurity interaction are
neglected or not is irrelevant to the dynamics in this case, and this is
reflected in the conservation of the positive semi-definiteness of the
electronic density matrix.
Nevertheless, the relevance of these correlations for the computation of the
observables grows as the number of available total spin states increases.
This is clearly exemplified in Fig.\,\ref{fig:fig_2}b by the abrupt change in
the relative difference between the approximate and the exact magnetization of
the impurities when $\beta=\pi/2$.

%%%%%%%%%%%%%%%%%%%%%%%%%%%%%%%%%%%%%%%%%%%%%%%%%%%%%%%%%%%%%%%%%%%%%%%%%%%%%%%
%%%%%%%%%%%%%%%%%%%%%%%%%%%%%%%%%%%%%%%%%%%%%%%%%%%%%%%%%%%%%%%%%%%%%%%%%%%%%%%

\section{Concluding remarks}
\label{sec:conclusions}

In this work we studied the quantum dynamics of a quasi-one-dimensional DMS 
quantum ring. 
The focus was on testing the methodological difficulties that appear
when employing the Heisenberg equations of motion to calculate
the dynamics of the electron and impurities' density matrices.
Following a standard scheme for DMS in the bulk, we truncated the infinite
hierarchy of equations by neglecting all direct impurity-impurity correlations
and reducing the indirect electron-mediated interaction to its mean-field terms.
Through this approach we obtained an approximate and numerically tractable set
of equations that goes beyond traditional mean-field approximations of the full
Hamiltonian.

In order to study the features and limitations of the truncation scheme, we
considered a small system of one electron and few magnetic impurities
initially in a pure state.
We integrated the exact time-dependent Schr\"odinger equation and used it to
compute the impurities' magnetization and the population of the electronic
states.
These results set a benchmark that allowed us to assess the accuracy of the
truncated set of equations.
We found that neglecting the indirect impurity-impurity correlations altogether
does not break the fundamental symmetries of the system, but nevertheless leads
to a non strictly Hamiltonian time evolution.
For different initial states (with and without entanglement between the
impurities) and a variety of initial configurations, we found that the energy,
the total SAM, the number of particles, and the hermiticity of the density
matrices are conserved to numerical precision.
The conservation of the number of particles (i.e., the traces of the electronic
and the $\langle P^I\rangle$ density matrices) is indicative that errors in the
populations are, up to numerical precision, exactly compensated at each time
step.
However, for some populations we observed a small but negative drift that leads
them to take on negative values which could not be ascribed to numerical error.
The positive semi-definiteness of the density matrices is therefore not
conserved throughout the time range studied.
In fact, in most cases it breaks right at the first time step.
From a theoretical point of view, this problem is rather serious and must be
addressed before using the truncated set of equations to study the physics of a
DMS QR in depth, particularly when no exact solution is at hand.
In practice, however, we saw that the approximation yields a remarkably accurate
estimation of an observable like the impurities' magnetization in the same time
range.
We conclude that, under certain conditions, the truncation scheme can still be
applied to study the dynamics of some physical quantities, particularly those
that are not too sensitive to errors in the populations, in time scales longer
than those of traditional time-dependent perturbation theory.

Finally, we mention that the problem of guaranteeing the positive semidefiniteness of a truncated or approximated density matrix has been studied in other areas of many-body physics.
A case in point is the field of atomic and molecular physics and the theory of reduced density matrices (see Ref.\ [\onlinecite{maz}]), which provides methods for computing physical properties of systems with many interacting electrons using only low-order density matrices (that is, density matrices involving few electronic creation and annihilitation operators).
It is known that such methods can sometimes yield density matrices that are not positive semidefinite and need to be corrected. This is in fact possible, but the procedure for correcting (or “purifying”) one particular density matrix in general requires imposing conditions on other density matrices of lower order as well (see Ref.\ [\onlinecite{alc}] and references therein).

This is also the case for the problem presented here, as any further approximation carried out on any of the density matrices would require guaranteeing also the conservation of the energy and total SAM, without breaking the symmetries of the system, which, as we mentioned, are respected by the original truncation scheme.
Such constraints couple all density matrices.
Some particular instances of this problem may be tackled using the techniques of semidefinite programming or the theory of convex optimization, for example to replace each density matrix with its optimal projection in the space of positive semidefinite ones that satisfy the required constraints. 
However, such a complex optimization problem would have to be solved after each numerical integration step, since positive semidefiniteness is not conserved, and even for a small number of impurities its computational cost may be prohibitively high.
Furthermore, such mathematical approaches do not necessarily follow
meaningful physical criteria that one may wish to enforce.
A solution that tackles the core of the physical problem directly 
in the truncation scheme itself would therefore be more desirable.
This work aims to contribute to the search of such a solution 
by singling out a serious drawback present in the conventional 
truncation scheme of the hierarchy of dynamical equations of
motion of the density matrices.

\section*{Acknowledgments}

We gratefully acknowledge financial support from Universidad de Buenos Aires
(UBACyT 2018, 20020170100711BA), CONICET (PIP 11220200100568CO), and ANPCyT
(PICT-2020-SERIEA-01082).

\appendix
\section{}
\label{sec:appendix_PPp}

The Heisenberg equations for the quantities $\langle P^{I}P^{I'}\rangle$ involve
only commutators of the form $[P^{I}P^{I'}, P^{I''}]$ which again yield terms
proportional to $P^{I}P^{I'}$.
As a consequence, the time evolution of each $\langle P^{I}P^{I'}\rangle$
depends only on four-point density matrices
$\langle c^{\dag}_{\lambda_1}c_{\lambda_2} P^{I}P^{I'}\rangle$.
If, as in the expression for $Q$, these four-point matrices are expanded
according to Eq. \eqref{eq:ccPP_expansion} and all factors
$\delta\langle P^{I}P^{I'}\rangle$ are expressed in terms of their correlated
and mean-field parts, it follows that dropping the term
$\delta\langle c^{\dag}_{\lambda_1} c_{\lambda_2} P^{I} P^{I'}\rangle$ suffices
to close the set of equations at the three-point level.
It is therefore possible to add the Heisenberg equations for the quantities
$\langle P^{I}P^{I'}\rangle$ when $I\neq I'$ to the original set containing
Eqs. \eqref{eq:cc_P_P}, \eqref{eq:cc_P_cc} and \eqref{eq:ccP} while keeping it
closed at the three-point level and without introducing further approximations.

\section{}
\label{sec:appendix_polygon}

Let us assume that the impurities are located at the vertices of an $N$-sided
regular polygon and consider a rotation $R$ in the dihedral group $D_{N}$.
The operation $RHR^{\dag}$ leaves the operators $L_z^2$ and
$\mathbf{S}_I\cdot\mathbf{s}$ invariant, but shifts the arguments of the delta
function operators by an integer multiple of $2\pi/N$.
This translation along the ring is a cyclic permutation that relocates the
impurities to different vertices on the same polygon, because it maps each
parameter $\varphi_I$ to some other $\varphi_{I'}$  (modulo
$2\pi$).
The rotated Hamiltonian can also be obtained by permuting the operators
$\mathbf{S}_I$ instead of shifting the delta potentials.
That is, the operation $RHR^{\dag}$ is equivalent to
$\hat{O}_{R}H\hat{O}_{R}^{\dag}$, for some $\hat{O}_R$ that relabels the
$\mathbf{S}_I$ without affecting the parameters $\varphi_I$.
The operator $\hat{O}_R$ can be written as a product of pairwise permutations
$\hat{O}_{II'} \doteq \sum_{n_1n_2}P^{I}_{n_1n_2}P^{I'}_{n_2n_1},$ that
interchanges two impurities in $H_{sd}$ by swapping the indices $I$ and $I'$
($I\neq I'$) of the operators $\mathbf{S}_{I}$ and $\mathbf{S}_{I'}$.
Notice that $\hat{O}_{II'}^{-1} = \hat{O}_{I'I} = \hat{O}_{II'}^{\dag} =
\hat{O}_{II'}$ as required.

Let us decompose the rotation $R$ into a product of two rotations,
$R=R_{\text{OAM}}R_{\text{SAM}}$, one acting only on the electron's OAM and the
other acting on its spin and the impurities' SAM, and consider the ket
$|\psi\rangle = |m\sigma\rangle|\text{Mn}\rangle\cdots|\text{Mn}\rangle$, where
$|m\sigma\rangle$ is an eigenstate of the operators $L_z$ and $s_z$ with
eigenvalues $m$ and $\sigma$ respectively, and $|\text{Mn}\rangle$ an arbitrary
single-impurity state that is repeated $N$ times in the product.
Notice that $|\psi\rangle$ is an eigenstate of $R_{\text{OAM}}$ and of $O_{R}$
for any rotation $R$, since swapping any pair of $|\text{Mn}\rangle$ factors in
$|\psi\rangle$ does not change the latter.
Calling $\mathcal{U}(t)$ the time-evolution operator and $S_z^{I}$ the
$z$-component of $\mathbf{S}_I$ in the Schr\"odinger picture, we write
\begin{equation}
  \begin{aligned}
    \langle\psi|\mathcal{U}^{\dag} S_z^{I}\mathcal{U}|\psi\rangle
    &= \langle\psi|
      \hat{O}_{R}^{\dag} R \mathcal{U}^{\dag}  R^{\dag}\hat{O}_{R}
      S_z^{I}
      \hat{O}_{R}^{\dag} R \mathcal{U}  R^{\dag} \hat{O}_{R}
      |\psi\rangle = \langle\psi| R_{\text{SAM}}^{\dag}  \mathcal{U}^{\dag}
      S_z^{I'} \mathcal{U} R_{\text{SAM}}|\psi\rangle \\
    &= \langle\psi| R_{\text{SAM}}^{\dag}  \mathcal{U}^{\dag}
      S_z^{I'} \mathcal{U} R_{\text{SAM}} |\psi\rangle = \langle\psi|
      \mathcal{U}^{\dag} S_z^{I'} \mathcal{U}|\psi\rangle
  \end{aligned}
\end{equation}
The second equality on the right-hand side follows from the equalities
$\hat{O}_{R}^{\dag} S_z^{I} \hat{O}_{R} = S_z^{I'}$ for some $I'\neq I$, and
$[S_z^I, R]=0$ for all $R$.
The fourth equality follows instead from the invariance of $H$ with respect to
spin-only rotations; that is, $[R_{\text{SAM}},H]=0$.

\bibliography{references.bib}

\begin{thebibliography}{44}%
\makeatletter
\providecommand \@ifxundefined [1]{%
 \@ifx{#1\undefined}
}%
\providecommand \@ifnum [1]{%
 \ifnum #1\expandafter \@firstoftwo
 \else \expandafter \@secondoftwo
 \fi
}%
\providecommand \@ifx [1]{%
 \ifx #1\expandafter \@firstoftwo
 \else \expandafter \@secondoftwo
 \fi
}%
\providecommand \natexlab [1]{#1}%
\providecommand \enquote  [1]{``#1''}%
\providecommand \bibnamefont  [1]{#1}%
\providecommand \bibfnamefont [1]{#1}%
\providecommand \citenamefont [1]{#1}%
\providecommand \href@noop [0]{\@secondoftwo}%
\providecommand \href [0]{\begingroup \@sanitize@url \@href}%
\providecommand \@href[1]{\@@startlink{#1}\@@href}%
\providecommand \@@href[1]{\endgroup#1\@@endlink}%
\providecommand \@sanitize@url [0]{\catcode `\\12\catcode `\$12\catcode
  `\&12\catcode `\#12\catcode `\^12\catcode `\_12\catcode `\%12\relax}%
\providecommand \@@startlink[1]{}%
\providecommand \@@endlink[0]{}%
\providecommand \url  [0]{\begingroup\@sanitize@url \@url }%
\providecommand \@url [1]{\endgroup\@href {#1}{\urlprefix }}%
\providecommand \urlprefix  [0]{URL }%
\providecommand \Eprint [0]{\href }%
\providecommand \doibase [0]{http://dx.doi.org/}%
\providecommand \selectlanguage [0]{\@gobble}%
\providecommand \bibinfo  [0]{\@secondoftwo}%
\providecommand \bibfield  [0]{\@secondoftwo}%
\providecommand \translation [1]{[#1]}%
\providecommand \BibitemOpen [0]{}%
\providecommand \bibitemStop [0]{}%
\providecommand \bibitemNoStop [0]{.\EOS\space}%
\providecommand \EOS [0]{\spacefactor3000\relax}%
\providecommand \BibitemShut  [1]{\csname bibitem#1\endcsname}%
\let\auto@bib@innerbib\@empty
%</preamble>
\bibitem [{\citenamefont {Thurn}\ and\ \citenamefont {Axt}(2012)}]{thu-axt}%
  \BibitemOpen
  \bibfield  {author} {\bibinfo {author} {\bibfnamefont {C.}~\bibnamefont
  {Thurn}}\ and\ \bibinfo {author} {\bibfnamefont {V.~M.}\ \bibnamefont
  {Axt}},\ }\href {\doibase 10.1103/PhysRevB.85.165203} {\bibfield  {journal}
  {\bibinfo  {journal} {Phys. Rev. B}\ }\textbf {\bibinfo {volume} {85}},\
  \bibinfo {pages} {165203} (\bibinfo {year} {2012})}\BibitemShut {NoStop}%
\bibitem [{\citenamefont {Aarts}\ \emph {et~al.}(2000)\citenamefont {Aarts},
  \citenamefont {Bonini},\ and\ \citenamefont {Wetterich}}]{aar-bon-wet}%
  \BibitemOpen
  \bibfield  {author} {\bibinfo {author} {\bibfnamefont {G.}~\bibnamefont
  {Aarts}}, \bibinfo {author} {\bibfnamefont {G.~F.}\ \bibnamefont {Bonini}}, \
  and\ \bibinfo {author} {\bibfnamefont {C.}~\bibnamefont {Wetterich}},\ }\href
  {\doibase 10.1103/PhysRevD.63.025012} {\bibfield  {journal} {\bibinfo
  {journal} {Phys. Rev. D}\ }\textbf {\bibinfo {volume} {63}},\ \bibinfo
  {pages} {025012} (\bibinfo {year} {2000})}\BibitemShut {NoStop}%
\bibitem [{\citenamefont {Leymann}\ \emph {et~al.}(2014)\citenamefont
  {Leymann}, \citenamefont {Foerster},\ and\ \citenamefont
  {Wiersig}}]{ley-foe-wie}%
  \BibitemOpen
  \bibfield  {author} {\bibinfo {author} {\bibfnamefont {H.~A.~M.}\
  \bibnamefont {Leymann}}, \bibinfo {author} {\bibfnamefont {A.}~\bibnamefont
  {Foerster}}, \ and\ \bibinfo {author} {\bibfnamefont {J.}~\bibnamefont
  {Wiersig}},\ }\href {\doibase 10.1103/PhysRevB.89.085308} {\bibfield
  {journal} {\bibinfo  {journal} {Phys. Rev. B}\ }\textbf {\bibinfo {volume}
  {89}},\ \bibinfo {pages} {085308} (\bibinfo {year} {2014})}\BibitemShut
  {NoStop}%
\bibitem [{\citenamefont {Xu}\ and\ \citenamefont {Yan}(2007)}]{xu-yan}%
  \BibitemOpen
  \bibfield  {author} {\bibinfo {author} {\bibfnamefont {R.-X.}\ \bibnamefont
  {Xu}}\ and\ \bibinfo {author} {\bibfnamefont {Y.-J.}\ \bibnamefont {Yan}},\
  }\href {\doibase 10.1103/PhysRevE.75.031107} {\bibfield  {journal} {\bibinfo
  {journal} {Phys. Rev. E}\ }\textbf {\bibinfo {volume} {75}},\ \bibinfo
  {pages} {031107} (\bibinfo {year} {2007})}\BibitemShut {NoStop}%
\bibitem [{\citenamefont {Kubo}(1962)}]{kub}%
  \BibitemOpen
  \bibfield  {author} {\bibinfo {author} {\bibfnamefont {R.}~\bibnamefont
  {Kubo}},\ }\href {\doibase 10.1143/JPSJ.17.1100} {\bibfield  {journal}
  {\bibinfo  {journal} {J. Phys. Soc. Jpn.}\ }\textbf {\bibinfo {volume}
  {17}},\ \bibinfo {pages} {1100} (\bibinfo {year} {1962})}\BibitemShut
  {NoStop}%
\bibitem [{\citenamefont {Axt}\ and\ \citenamefont {Stahl}(1994)}]{axt-sta}%
  \BibitemOpen
  \bibfield  {author} {\bibinfo {author} {\bibfnamefont {V.~M.}\ \bibnamefont
  {Axt}}\ and\ \bibinfo {author} {\bibfnamefont {A.}~\bibnamefont {Stahl}},\
  }\href {\doibase 10.1007/BF01316963} {\bibfield  {journal} {\bibinfo
  {journal} {Zeitschrift f{\"u}r Physik B Condensed Matter}\ }\textbf {\bibinfo
  {volume} {93}},\ \bibinfo {pages} {195} (\bibinfo {year} {1994})}\BibitemShut
  {NoStop}%
\bibitem [{\citenamefont {Rossi}\ and\ \citenamefont {Kuhn}(2002)}]{ros-kuh}%
  \BibitemOpen
  \bibfield  {author} {\bibinfo {author} {\bibfnamefont {F.}~\bibnamefont
  {Rossi}}\ and\ \bibinfo {author} {\bibfnamefont {T.}~\bibnamefont {Kuhn}},\
  }\href {\doibase 10.1103/RevModPhys.74.895} {\bibfield  {journal} {\bibinfo
  {journal} {Rev. Mod. Phys.}\ }\textbf {\bibinfo {volume} {74}},\ \bibinfo
  {pages} {895} (\bibinfo {year} {2002})}\BibitemShut {NoStop}%
\bibitem [{\citenamefont {Khitrova}\ \emph {et~al.}(1999)\citenamefont
  {Khitrova}, \citenamefont {Gibbs}, \citenamefont {Jahnke}, \citenamefont
  {Kira},\ and\ \citenamefont {Koch}}]{khi-gib-jan}%
  \BibitemOpen
  \bibfield  {author} {\bibinfo {author} {\bibfnamefont {G.}~\bibnamefont
  {Khitrova}}, \bibinfo {author} {\bibfnamefont {H.~M.}\ \bibnamefont {Gibbs}},
  \bibinfo {author} {\bibfnamefont {F.}~\bibnamefont {Jahnke}}, \bibinfo
  {author} {\bibfnamefont {M.}~\bibnamefont {Kira}}, \ and\ \bibinfo {author}
  {\bibfnamefont {S.~W.}\ \bibnamefont {Koch}},\ }\href {\doibase
  10.1103/RevModPhys.71.1591} {\bibfield  {journal} {\bibinfo  {journal} {Rev.
  Mod. Phys.}\ }\textbf {\bibinfo {volume} {71}},\ \bibinfo {pages} {1591}
  (\bibinfo {year} {1999})}\BibitemShut {NoStop}%
\bibitem [{\citenamefont {Lindberg}\ \emph {et~al.}(1994)\citenamefont
  {Lindberg}, \citenamefont {Hu}, \citenamefont {Binder},\ and\ \citenamefont
  {Koch}}]{lin-hu-bin}%
  \BibitemOpen
  \bibfield  {author} {\bibinfo {author} {\bibfnamefont {M.}~\bibnamefont
  {Lindberg}}, \bibinfo {author} {\bibfnamefont {Y.~Z.}\ \bibnamefont {Hu}},
  \bibinfo {author} {\bibfnamefont {R.}~\bibnamefont {Binder}}, \ and\ \bibinfo
  {author} {\bibfnamefont {S.~W.}\ \bibnamefont {Koch}},\ }\href {\doibase
  10.1103/PhysRevB.50.18060} {\bibfield  {journal} {\bibinfo  {journal} {Phys.
  Rev. B}\ }\textbf {\bibinfo {volume} {50}},\ \bibinfo {pages} {18060}
  (\bibinfo {year} {1994})}\BibitemShut {NoStop}%
\bibitem [{\citenamefont {Ungar}\ \emph {et~al.}(2017)\citenamefont {Ungar},
  \citenamefont {Cygorek},\ and\ \citenamefont {Axt}}]{ung-cyg-axt}%
  \BibitemOpen
  \bibfield  {author} {\bibinfo {author} {\bibfnamefont {F.}~\bibnamefont
  {Ungar}}, \bibinfo {author} {\bibfnamefont {M.}~\bibnamefont {Cygorek}}, \
  and\ \bibinfo {author} {\bibfnamefont {V.~M.}\ \bibnamefont {Axt}},\ }\href
  {\doibase 10.1103/PhysRevB.95.245203} {\bibfield  {journal} {\bibinfo
  {journal} {Phys. Rev. B}\ }\textbf {\bibinfo {volume} {95}},\ \bibinfo
  {pages} {245203} (\bibinfo {year} {2017})}\BibitemShut {NoStop}%
\bibitem [{\citenamefont {Ungar}\ \emph {et~al.}(2018)\citenamefont {Ungar},
  \citenamefont {Cygorek},\ and\ \citenamefont {Axt}}]{ung-cyg-axt-2}%
  \BibitemOpen
  \bibfield  {author} {\bibinfo {author} {\bibfnamefont {F.}~\bibnamefont
  {Ungar}}, \bibinfo {author} {\bibfnamefont {M.}~\bibnamefont {Cygorek}}, \
  and\ \bibinfo {author} {\bibfnamefont {V.~M.}\ \bibnamefont {Axt}},\ }\href
  {\doibase 10.1103/PhysRevB.97.045210} {\bibfield  {journal} {\bibinfo
  {journal} {Phys. Rev. B}\ }\textbf {\bibinfo {volume} {97}},\ \bibinfo
  {pages} {045210} (\bibinfo {year} {2018})}\BibitemShut {NoStop}%
\bibitem [{\citenamefont {Kacman}(2001)}]{kac}%
  \BibitemOpen
  \bibfield  {author} {\bibinfo {author} {\bibfnamefont {P.}~\bibnamefont
  {Kacman}},\ }\href {\doibase 10.1088/0268-1242/16/4/201} {\bibfield
  {journal} {\bibinfo  {journal} {Semiconductor Science and Technology}\
  }\textbf {\bibinfo {volume} {16}},\ \bibinfo {pages} {R25} (\bibinfo {year}
  {2001})}\BibitemShut {NoStop}%
\bibitem [{\citenamefont {Blinowski}\ and\ \citenamefont
  {Kacman}(2003)}]{bli-kac}%
  \BibitemOpen
  \bibfield  {author} {\bibinfo {author} {\bibfnamefont {J.}~\bibnamefont
  {Blinowski}}\ and\ \bibinfo {author} {\bibfnamefont {P.}~\bibnamefont
  {Kacman}},\ }\href {\doibase 10.1103/PhysRevB.67.121204} {\bibfield
  {journal} {\bibinfo  {journal} {Phys. Rev. B}\ }\textbf {\bibinfo {volume}
  {67}},\ \bibinfo {pages} {121204(R)} (\bibinfo {year} {2003})}\BibitemShut
  {NoStop}%
\bibitem [{\citenamefont {Morandi}\ \emph {et~al.}(2009)\citenamefont
  {Morandi}, \citenamefont {Hervieux},\ and\ \citenamefont {Manfredi}}]{mor}%
  \BibitemOpen
  \bibfield  {author} {\bibinfo {author} {\bibfnamefont {O.}~\bibnamefont
  {Morandi}}, \bibinfo {author} {\bibfnamefont {P.-A.}\ \bibnamefont
  {Hervieux}}, \ and\ \bibinfo {author} {\bibfnamefont {G.}~\bibnamefont
  {Manfredi}},\ }\href {\doibase 10.1088/1367-2630/11/7/073010} {\bibfield
  {journal} {\bibinfo  {journal} {New Journal of Physics}\ }\textbf {\bibinfo
  {volume} {11}},\ \bibinfo {pages} {073010} (\bibinfo {year}
  {2009})}\BibitemShut {NoStop}%
\bibitem [{\citenamefont {Chang}\ \emph {et~al.}(2004)\citenamefont {Chang},
  \citenamefont {Li}, \citenamefont {Xia},\ and\ \citenamefont
  {Peeters}}]{cha-li-xia}%
  \BibitemOpen
  \bibfield  {author} {\bibinfo {author} {\bibfnamefont {K.}~\bibnamefont
  {Chang}}, \bibinfo {author} {\bibfnamefont {S.~S.}\ \bibnamefont {Li}},
  \bibinfo {author} {\bibfnamefont {J.~B.}\ \bibnamefont {Xia}}, \ and\
  \bibinfo {author} {\bibfnamefont {F.~M.}\ \bibnamefont {Peeters}},\ }\href
  {\doibase 10.1103/PhysRevB.69.235203} {\bibfield  {journal} {\bibinfo
  {journal} {Phys. Rev. B}\ }\textbf {\bibinfo {volume} {69}},\ \bibinfo
  {pages} {235203} (\bibinfo {year} {2004})}\BibitemShut {NoStop}%
\bibitem [{\citenamefont {Wu}\ \emph {et~al.}(2010)\citenamefont {Wu},
  \citenamefont {Jiang},\ and\ \citenamefont {Weng}}]{wu-jia-wen}%
  \BibitemOpen
  \bibfield  {author} {\bibinfo {author} {\bibfnamefont {M.}~\bibnamefont
  {Wu}}, \bibinfo {author} {\bibfnamefont {J.}~\bibnamefont {Jiang}}, \ and\
  \bibinfo {author} {\bibfnamefont {M.}~\bibnamefont {Weng}},\ }\href {\doibase
  https://doi.org/10.1016/j.physrep.2010.04.002} {\bibfield  {journal}
  {\bibinfo  {journal} {Physics Reports}\ }\textbf {\bibinfo {volume} {493}},\
  \bibinfo {pages} {61 } (\bibinfo {year} {2010})}\BibitemShut {NoStop}%
\bibitem [{\citenamefont {Ma}(2013)}]{ma}%
  \BibitemOpen
  \bibfield  {author} {\bibinfo {author} {\bibfnamefont {X.}~\bibnamefont
  {Ma}},\ }\href {\doibase 10.1007/s10853-012-6985-y} {\bibfield  {journal}
  {\bibinfo  {journal} {Journal of Materials Science}\ }\textbf {\bibinfo
  {volume} {48}},\ \bibinfo {pages} {2111} (\bibinfo {year}
  {2013})}\BibitemShut {NoStop}%
\bibitem [{\citenamefont {Krainov}\ \emph {et~al.}(2017)\citenamefont
  {Krainov}, \citenamefont {Vladimirova}, \citenamefont {Scalbert},
  \citenamefont {L\"ahderanta}, \citenamefont {Dmitriev},\ and\ \citenamefont
  {Averkiev}}]{kra-vla-sca}%
  \BibitemOpen
  \bibfield  {author} {\bibinfo {author} {\bibfnamefont {I.~V.}\ \bibnamefont
  {Krainov}}, \bibinfo {author} {\bibfnamefont {M.}~\bibnamefont
  {Vladimirova}}, \bibinfo {author} {\bibfnamefont {D.}~\bibnamefont
  {Scalbert}}, \bibinfo {author} {\bibfnamefont {E.}~\bibnamefont
  {L\"ahderanta}}, \bibinfo {author} {\bibfnamefont {A.~P.}\ \bibnamefont
  {Dmitriev}}, \ and\ \bibinfo {author} {\bibfnamefont {N.~S.}\ \bibnamefont
  {Averkiev}},\ }\href {\doibase 10.1103/PhysRevB.96.165304} {\bibfield
  {journal} {\bibinfo  {journal} {Phys. Rev. B}\ }\textbf {\bibinfo {volume}
  {96}},\ \bibinfo {pages} {165304} (\bibinfo {year} {2017})}\BibitemShut
  {NoStop}%
\bibitem [{\citenamefont {Ungar}\ \emph {et~al.}(2019)\citenamefont {Ungar},
  \citenamefont {Cygorek},\ and\ \citenamefont {Axt}}]{ung-cyg-axt-4}%
  \BibitemOpen
  \bibfield  {author} {\bibinfo {author} {\bibfnamefont {F.}~\bibnamefont
  {Ungar}}, \bibinfo {author} {\bibfnamefont {M.}~\bibnamefont {Cygorek}}, \
  and\ \bibinfo {author} {\bibfnamefont {V.~M.}\ \bibnamefont {Axt}},\ }\href
  {\doibase 10.1103/PhysRevB.99.195309} {\bibfield  {journal} {\bibinfo
  {journal} {Phys. Rev. B}\ }\textbf {\bibinfo {volume} {99}},\ \bibinfo
  {pages} {195309} (\bibinfo {year} {2019})}\BibitemShut {NoStop}%
\bibitem [{\citenamefont {Viefers}\ \emph {et~al.}(2004)\citenamefont
  {Viefers}, \citenamefont {Koskinen}, \citenamefont {{Singha Deo}},\ and\
  \citenamefont {Manninen}}]{vie-kos-sin}%
  \BibitemOpen
  \bibfield  {author} {\bibinfo {author} {\bibfnamefont {S.}~\bibnamefont
  {Viefers}}, \bibinfo {author} {\bibfnamefont {P.}~\bibnamefont {Koskinen}},
  \bibinfo {author} {\bibfnamefont {P.}~\bibnamefont {{Singha Deo}}}, \ and\
  \bibinfo {author} {\bibfnamefont {M.}~\bibnamefont {Manninen}},\ }\href
  {\doibase https://doi.org/10.1016/j.physe.2003.08.076} {\bibfield  {journal}
  {\bibinfo  {journal} {Physica E: Low-dimensional Systems and Nanostructures}\
  }\textbf {\bibinfo {volume} {21}},\ \bibinfo {pages} {1} (\bibinfo {year}
  {2004})}\BibitemShut {NoStop}%
\bibitem [{\citenamefont {Dietl}(2010)}]{die}%
  \BibitemOpen
  \bibfield  {author} {\bibinfo {author} {\bibfnamefont {T.}~\bibnamefont
  {Dietl}},\ }\href {\doibase 10.1038/nmat2898} {\bibfield  {journal} {\bibinfo
   {journal} {Nature Materials}\ }\textbf {\bibinfo {volume} {9}},\ \bibinfo
  {pages} {965} (\bibinfo {year} {2010})}\BibitemShut {NoStop}%
\bibitem [{\citenamefont {Dietl}\ and\ \citenamefont {Ohno}(2014)}]{die-ohn}%
  \BibitemOpen
  \bibfield  {author} {\bibinfo {author} {\bibfnamefont {T.}~\bibnamefont
  {Dietl}}\ and\ \bibinfo {author} {\bibfnamefont {H.}~\bibnamefont {Ohno}},\
  }\href {\doibase 10.1103/RevModPhys.86.187} {\bibfield  {journal} {\bibinfo
  {journal} {Rev. Mod. Phys.}\ }\textbf {\bibinfo {volume} {86}},\ \bibinfo
  {pages} {187} (\bibinfo {year} {2014})}\BibitemShut {NoStop}%
\bibitem [{\citenamefont {Frustaglia}\ and\ \citenamefont
  {Richter}(2004)}]{fru-ric}%
  \BibitemOpen
  \bibfield  {author} {\bibinfo {author} {\bibfnamefont {D.}~\bibnamefont
  {Frustaglia}}\ and\ \bibinfo {author} {\bibfnamefont {K.}~\bibnamefont
  {Richter}},\ }\href {\doibase 10.1103/PhysRevB.69.235310} {\bibfield
  {journal} {\bibinfo  {journal} {Phys. Rev. B}\ }\textbf {\bibinfo {volume}
  {69}},\ \bibinfo {pages} {235310} (\bibinfo {year} {2004})}\BibitemShut
  {NoStop}%
\bibitem [{\citenamefont {Yakovlev}\ and\ \citenamefont
  {Merkulov}(2010)}]{yak-mer}%
  \BibitemOpen
  \bibfield  {author} {\bibinfo {author} {\bibfnamefont {D.~R.}\ \bibnamefont
  {Yakovlev}}\ and\ \bibinfo {author} {\bibfnamefont {I.~A.}\ \bibnamefont
  {Merkulov}},\ }\enquote {\bibinfo {title} {Spin and energy transfer between
  carriers, magnetic ions, and lattice},}\ in\ \href {\doibase
  10.1007/978-3-642-15856-8_8} {\emph {\bibinfo {booktitle} {Introduction to
  the Physics of Diluted Magnetic Semiconductors}}},\ \bibinfo {editor} {edited
  by\ \bibinfo {editor} {\bibfnamefont {J.~A.}\ \bibnamefont {Gaj}}\ and\
  \bibinfo {editor} {\bibfnamefont {J.}~\bibnamefont {Kossut}}}\ (\bibinfo
  {publisher} {Springer Berlin Heidelberg},\ \bibinfo {address} {Berlin,
  Heidelberg},\ \bibinfo {year} {2010})\ pp.\ \bibinfo {pages}
  {263--303}\BibitemShut {NoStop}%
\bibitem [{\citenamefont {Kondo}(1964)}]{kon}%
  \BibitemOpen
  \bibfield  {author} {\bibinfo {author} {\bibfnamefont {J.}~\bibnamefont
  {Kondo}},\ }\href {\doibase 10.1143/PTP.32.37} {\bibfield  {journal}
  {\bibinfo  {journal} {Progress of Theoretical Physics}\ }\textbf {\bibinfo
  {volume} {32}},\ \bibinfo {pages} {37} (\bibinfo {year} {1964})}\BibitemShut
  {NoStop}%
\bibitem [{\citenamefont {Lia}\ and\ \citenamefont
  {Tamborenea}(2021)}]{lia-tam}%
  \BibitemOpen
  \bibfield  {author} {\bibinfo {author} {\bibfnamefont {J.~M.}\ \bibnamefont
  {Lia}}\ and\ \bibinfo {author} {\bibfnamefont {P.~I.}\ \bibnamefont
  {Tamborenea}},\ }\href {\doibase https://doi.org/10.1016/j.physe.2020.114419}
  {\bibfield  {journal} {\bibinfo  {journal} {Physica E: Low-dimensional
  Systems and Nanostructures}\ }\textbf {\bibinfo {volume} {126}},\ \bibinfo
  {pages} {114419} (\bibinfo {year} {2021})}\BibitemShut {NoStop}%
\bibitem [{\citenamefont {Lia}\ \emph {et~al.}(2022)\citenamefont {Lia},
  \citenamefont {Tamborenea}, \citenamefont {Cygorek},\ and\ \citenamefont
  {Axt}}]{lia-tam-cyg}%
  \BibitemOpen
  \bibfield  {author} {\bibinfo {author} {\bibfnamefont {J.~M.}\ \bibnamefont
  {Lia}}, \bibinfo {author} {\bibfnamefont {P.~I.}\ \bibnamefont {Tamborenea}},
  \bibinfo {author} {\bibfnamefont {M.}~\bibnamefont {Cygorek}}, \ and\
  \bibinfo {author} {\bibfnamefont {V.~M.}\ \bibnamefont {Axt}},\ }\href
  {\doibase 10.1103/PhysRevB.105.115426} {\bibfield  {journal} {\bibinfo
  {journal} {Phys. Rev. B}\ }\textbf {\bibinfo {volume} {105}},\ \bibinfo
  {pages} {115426} (\bibinfo {year} {2022})}\BibitemShut {NoStop}%
\bibitem [{\citenamefont {Qu}\ and\ \citenamefont {Hawrylak}(2005)}]{qu-haw}%
  \BibitemOpen
  \bibfield  {author} {\bibinfo {author} {\bibfnamefont {F.}~\bibnamefont
  {Qu}}\ and\ \bibinfo {author} {\bibfnamefont {P.}~\bibnamefont {Hawrylak}},\
  }\href {\doibase 10.1103/PhysRevLett.95.217206} {\bibfield  {journal}
  {\bibinfo  {journal} {Phys. Rev. Lett.}\ }\textbf {\bibinfo {volume} {95}},\
  \bibinfo {pages} {217206} (\bibinfo {year} {2005})}\BibitemShut {NoStop}%
\bibitem [{\citenamefont {Meijer}\ \emph {et~al.}(2002)\citenamefont {Meijer},
  \citenamefont {Morpurgo},\ and\ \citenamefont {Klapwijk}}]{mei-mor-klap}%
  \BibitemOpen
  \bibfield  {author} {\bibinfo {author} {\bibfnamefont {F.~E.}\ \bibnamefont
  {Meijer}}, \bibinfo {author} {\bibfnamefont {A.~F.}\ \bibnamefont
  {Morpurgo}}, \ and\ \bibinfo {author} {\bibfnamefont {T.~M.}\ \bibnamefont
  {Klapwijk}},\ }\href {\doibase 10.1103/PhysRevB.66.033107} {\bibfield
  {journal} {\bibinfo  {journal} {Phys. Rev. B}\ }\textbf {\bibinfo {volume}
  {66}},\ \bibinfo {pages} {033107} (\bibinfo {year} {2002})}\BibitemShut
  {NoStop}%
\bibitem [{\citenamefont {Lorke}\ \emph {et~al.}(2000)\citenamefont {Lorke},
  \citenamefont {Luyken}, \citenamefont {Govorov}, \citenamefont {Kotthaus},
  \citenamefont {Garcia},\ and\ \citenamefont {Petroff}}]{lor-joh-kot}%
  \BibitemOpen
  \bibfield  {author} {\bibinfo {author} {\bibfnamefont {A.}~\bibnamefont
  {Lorke}}, \bibinfo {author} {\bibfnamefont {R.~J.}\ \bibnamefont {Luyken}},
  \bibinfo {author} {\bibfnamefont {A.~O.}\ \bibnamefont {Govorov}}, \bibinfo
  {author} {\bibfnamefont {J.~P.}\ \bibnamefont {Kotthaus}}, \bibinfo {author}
  {\bibfnamefont {J.~M.}\ \bibnamefont {Garcia}}, \ and\ \bibinfo {author}
  {\bibfnamefont {P.~M.}\ \bibnamefont {Petroff}},\ }\href {\doibase
  10.1103/PhysRevLett.84.2223} {\bibfield  {journal} {\bibinfo  {journal}
  {Phys. Rev. Lett.}\ }\textbf {\bibinfo {volume} {84}},\ \bibinfo {pages}
  {2223} (\bibinfo {year} {2000})}\BibitemShut {NoStop}%
\bibitem [{\citenamefont {Lin}\ \emph {et~al.}(2009)\citenamefont {Lin},
  \citenamefont {Lin}, \citenamefont {Ling}, \citenamefont {Fu}, \citenamefont
  {Chang}, \citenamefont {Lin},\ and\ \citenamefont {Lee}}]{lin-lin-ling}%
  \BibitemOpen
  \bibfield  {author} {\bibinfo {author} {\bibfnamefont {T.-C.}\ \bibnamefont
  {Lin}}, \bibinfo {author} {\bibfnamefont {C.-H.}\ \bibnamefont {Lin}},
  \bibinfo {author} {\bibfnamefont {H.-S.}\ \bibnamefont {Ling}}, \bibinfo
  {author} {\bibfnamefont {Y.-J.}\ \bibnamefont {Fu}}, \bibinfo {author}
  {\bibfnamefont {W.-H.}\ \bibnamefont {Chang}}, \bibinfo {author}
  {\bibfnamefont {S.-D.}\ \bibnamefont {Lin}}, \ and\ \bibinfo {author}
  {\bibfnamefont {C.-P.}\ \bibnamefont {Lee}},\ }\href {\doibase
  10.1103/PhysRevB.80.081304} {\bibfield  {journal} {\bibinfo  {journal} {Phys.
  Rev. B}\ }\textbf {\bibinfo {volume} {80}},\ \bibinfo {pages} {081304(R)}
  (\bibinfo {year} {2009})}\BibitemShut {NoStop}%
\bibitem [{\citenamefont {Sheng}\ and\ \citenamefont
  {Chang}(2007)}]{she-cha-1}%
  \BibitemOpen
  \bibfield  {author} {\bibinfo {author} {\bibfnamefont {J.~S.}\ \bibnamefont
  {Sheng}}\ and\ \bibinfo {author} {\bibfnamefont {K.}~\bibnamefont {Chang}},\
  }\href {\doibase 10.1088/0953-8984/20/02/025222} {\bibfield  {journal}
  {\bibinfo  {journal} {Journal of Physics: Condensed Matter}\ }\textbf
  {\bibinfo {volume} {20}},\ \bibinfo {pages} {025222} (\bibinfo {year}
  {2007})}\BibitemShut {NoStop}%
\bibitem [{\citenamefont {Chakraborty}\ and\ \citenamefont
  {Pietil\"ainen}(1994)}]{cha-pie}%
  \BibitemOpen
  \bibfield  {author} {\bibinfo {author} {\bibfnamefont {T.}~\bibnamefont
  {Chakraborty}}\ and\ \bibinfo {author} {\bibfnamefont {P.}~\bibnamefont
  {Pietil\"ainen}},\ }\href {\doibase 10.1103/PhysRevB.50.8460} {\bibfield
  {journal} {\bibinfo  {journal} {Phys. Rev. B}\ }\textbf {\bibinfo {volume}
  {50}},\ \bibinfo {pages} {8460} (\bibinfo {year} {1994})}\BibitemShut
  {NoStop}%
\bibitem [{\citenamefont {Shakouri}\ \emph {et~al.}(2012)\citenamefont
  {Shakouri}, \citenamefont {Szafran}, \citenamefont {Esmaeilzadeh},\ and\
  \citenamefont {Peeters}}]{sha-sza-esm}%
  \BibitemOpen
  \bibfield  {author} {\bibinfo {author} {\bibfnamefont {K.}~\bibnamefont
  {Shakouri}}, \bibinfo {author} {\bibfnamefont {B.}~\bibnamefont {Szafran}},
  \bibinfo {author} {\bibfnamefont {M.}~\bibnamefont {Esmaeilzadeh}}, \ and\
  \bibinfo {author} {\bibfnamefont {F.~M.}\ \bibnamefont {Peeters}},\ }\href
  {\doibase 10.1103/PhysRevB.85.165314} {\bibfield  {journal} {\bibinfo
  {journal} {Phys. Rev. B}\ }\textbf {\bibinfo {volume} {85}},\ \bibinfo
  {pages} {165314} (\bibinfo {year} {2012})}\BibitemShut {NoStop}%
\bibitem [{\citenamefont {Furdyna}(1988)}]{fur}%
  \BibitemOpen
  \bibfield  {author} {\bibinfo {author} {\bibfnamefont {J.~K.}\ \bibnamefont
  {Furdyna}},\ }\href {\doibase 10.1063/1.341700} {\bibfield  {journal}
  {\bibinfo  {journal} {Journal of Applied Physics}\ }\textbf {\bibinfo
  {volume} {64}},\ \bibinfo {pages} {R29} (\bibinfo {year} {1988})}\BibitemShut
  {NoStop}%
\bibitem [{\citenamefont {Quinteiro}\ and\ \citenamefont
  {Tamborenea}(2009)}]{qui-tam}%
  \BibitemOpen
  \bibfield  {author} {\bibinfo {author} {\bibfnamefont {G.~F.}\ \bibnamefont
  {Quinteiro}}\ and\ \bibinfo {author} {\bibfnamefont {P.~I.}\ \bibnamefont
  {Tamborenea}},\ }\href {\doibase 10.1103/PhysRevB.79.155450} {\bibfield
  {journal} {\bibinfo  {journal} {Phys. Rev. B}\ }\textbf {\bibinfo {volume}
  {79}},\ \bibinfo {pages} {155450} (\bibinfo {year} {2009})}\BibitemShut
  {NoStop}%
\bibitem [{\citenamefont {Quinteiro}\ \emph {et~al.}(2011)\citenamefont
  {Quinteiro}, \citenamefont {Tamborenea},\ and\ \citenamefont
  {Berakdar}}]{qui-tam-ber}%
  \BibitemOpen
  \bibfield  {author} {\bibinfo {author} {\bibfnamefont {G.~F.}\ \bibnamefont
  {Quinteiro}}, \bibinfo {author} {\bibfnamefont {P.~I.}\ \bibnamefont
  {Tamborenea}}, \ and\ \bibinfo {author} {\bibfnamefont {J.}~\bibnamefont
  {Berakdar}},\ }\href {\doibase 10.1364/OE.19.026733} {\bibfield  {journal}
  {\bibinfo  {journal} {Opt. Express}\ }\textbf {\bibinfo {volume} {19}},\
  \bibinfo {pages} {26733} (\bibinfo {year} {2011})}\BibitemShut {NoStop}%
\bibitem [{\citenamefont {Quinteiro~Rosen}\ \emph {et~al.}(2022)\citenamefont
  {Quinteiro~Rosen}, \citenamefont {Tamborenea},\ and\ \citenamefont
  {Kuhn}}]{qui-tam-kuh}%
  \BibitemOpen
  \bibfield  {author} {\bibinfo {author} {\bibfnamefont {G.~F.}\ \bibnamefont
  {Quinteiro~Rosen}}, \bibinfo {author} {\bibfnamefont {P.~I.}\ \bibnamefont
  {Tamborenea}}, \ and\ \bibinfo {author} {\bibfnamefont {T.}~\bibnamefont
  {Kuhn}},\ }\href {\doibase 10.1103/RevModPhys.94.035003} {\bibfield
  {journal} {\bibinfo  {journal} {Rev. Mod. Phys.}\ }\textbf {\bibinfo {volume}
  {94}},\ \bibinfo {pages} {035003} (\bibinfo {year} {2022})}\BibitemShut
  {NoStop}%
\bibitem [{\citenamefont {Mike}\ \emph {et~al.}(2018)\citenamefont {Mike},
  \citenamefont {Szab{\'o}},\ and\ \citenamefont {F{\"o}ldi}}]{mik-sza-foe}%
  \BibitemOpen
  \bibfield  {author} {\bibinfo {author} {\bibfnamefont {P.}~\bibnamefont
  {Mike}}, \bibinfo {author} {\bibfnamefont {L.~Z.}\ \bibnamefont {Szab{\'o}}},
  \ and\ \bibinfo {author} {\bibfnamefont {P.}~\bibnamefont {F{\"o}ldi}},\
  }\href {\doibase 10.1007/s10946-018-9741-1} {\bibfield  {journal} {\bibinfo
  {journal} {Journal of Russian Laser Research}\ }\textbf {\bibinfo {volume}
  {39}},\ \bibinfo {pages} {465} (\bibinfo {year} {2018})}\BibitemShut
  {NoStop}%
\bibitem [{\citenamefont {Kneip}\ \emph {et~al.}(2006)\citenamefont {Kneip},
  \citenamefont {Yakovlev}, \citenamefont {Bayer}, \citenamefont {Maksimov},
  \citenamefont {Tartakovskii}, \citenamefont {Keller}, \citenamefont {Ossau},
  \citenamefont {Molenkamp},\ and\ \citenamefont {Waag}}]{kne-yak-bay}%
  \BibitemOpen
  \bibfield  {author} {\bibinfo {author} {\bibfnamefont {M.~K.}\ \bibnamefont
  {Kneip}}, \bibinfo {author} {\bibfnamefont {D.~R.}\ \bibnamefont {Yakovlev}},
  \bibinfo {author} {\bibfnamefont {M.}~\bibnamefont {Bayer}}, \bibinfo
  {author} {\bibfnamefont {A.~A.}\ \bibnamefont {Maksimov}}, \bibinfo {author}
  {\bibfnamefont {I.~I.}\ \bibnamefont {Tartakovskii}}, \bibinfo {author}
  {\bibfnamefont {D.}~\bibnamefont {Keller}}, \bibinfo {author} {\bibfnamefont
  {W.}~\bibnamefont {Ossau}}, \bibinfo {author} {\bibfnamefont {L.~W.}\
  \bibnamefont {Molenkamp}}, \ and\ \bibinfo {author} {\bibfnamefont
  {A.}~\bibnamefont {Waag}},\ }\href {\doibase 10.1103/PhysRevB.73.035306}
  {\bibfield  {journal} {\bibinfo  {journal} {Phys. Rev. B}\ }\textbf {\bibinfo
  {volume} {73}},\ \bibinfo {pages} {035306} (\bibinfo {year}
  {2006})}\BibitemShut {NoStop}%
\bibitem [{\citenamefont {Dietl}\ \emph {et~al.}(1995)\citenamefont {Dietl},
  \citenamefont {Peyla}, \citenamefont {Grieshaber},\ and\ \citenamefont
  {Merle~d'Aubign\'e}}]{die-pey-gri}%
  \BibitemOpen
  \bibfield  {author} {\bibinfo {author} {\bibfnamefont {T.}~\bibnamefont
  {Dietl}}, \bibinfo {author} {\bibfnamefont {P.}~\bibnamefont {Peyla}},
  \bibinfo {author} {\bibfnamefont {W.}~\bibnamefont {Grieshaber}}, \ and\
  \bibinfo {author} {\bibfnamefont {Y.}~\bibnamefont {Merle~d'Aubign\'e}},\
  }\href {\doibase 10.1103/PhysRevLett.74.474} {\bibfield  {journal} {\bibinfo
  {journal} {Phys. Rev. Lett.}\ }\textbf {\bibinfo {volume} {74}},\ \bibinfo
  {pages} {474} (\bibinfo {year} {1995})}\BibitemShut {NoStop}%
\bibitem [{\citenamefont {K\"onig}\ \emph {et~al.}(2000)\citenamefont
  {K\"onig}, \citenamefont {Merkulov}, \citenamefont {Yakovlev}, \citenamefont
  {Ossau}, \citenamefont {Ryabchenko}, \citenamefont {Kutrowski}, \citenamefont
  {Wojtowicz}, \citenamefont {Karczewski},\ and\ \citenamefont
  {Kossut}}]{koe-mer-yak}%
  \BibitemOpen
  \bibfield  {author} {\bibinfo {author} {\bibfnamefont {B.}~\bibnamefont
  {K\"onig}}, \bibinfo {author} {\bibfnamefont {I.~A.}\ \bibnamefont
  {Merkulov}}, \bibinfo {author} {\bibfnamefont {D.~R.}\ \bibnamefont
  {Yakovlev}}, \bibinfo {author} {\bibfnamefont {W.}~\bibnamefont {Ossau}},
  \bibinfo {author} {\bibfnamefont {S.~M.}\ \bibnamefont {Ryabchenko}},
  \bibinfo {author} {\bibfnamefont {M.}~\bibnamefont {Kutrowski}}, \bibinfo
  {author} {\bibfnamefont {T.}~\bibnamefont {Wojtowicz}}, \bibinfo {author}
  {\bibfnamefont {G.}~\bibnamefont {Karczewski}}, \ and\ \bibinfo {author}
  {\bibfnamefont {J.}~\bibnamefont {Kossut}},\ }\href {\doibase
  10.1103/PhysRevB.61.16870} {\bibfield  {journal} {\bibinfo  {journal} {Phys.
  Rev. B}\ }\textbf {\bibinfo {volume} {61}},\ \bibinfo {pages} {16870}
  (\bibinfo {year} {2000})}\BibitemShut {NoStop}%
\bibitem [{\citenamefont {Mazziotti}(2007)}]{maz}%
  \BibitemOpen
  \bibinfo {editor} {\bibfnamefont {D.~A.}\ \bibnamefont {Mazziotti}},\ ed.,\
  \href {\doibase https://doi.org/10.1002/0470106603} {\emph {\bibinfo {title}
  {Reduced‐Density‐Matrix Mechanics: With Application to Many‐Electron
  Atoms and Molecules}}}\ (\bibinfo  {publisher} {John Wiley \& Sons, Ltd},\
  \bibinfo {year} {2007})\BibitemShut {NoStop}%
\bibitem [{\citenamefont {Alcoba}(2007)}]{alc}%
  \BibitemOpen
  \bibfield  {author} {\bibinfo {author} {\bibfnamefont {D.~R.}\ \bibnamefont
  {Alcoba}},\ }\enquote {\bibinfo {title} {Purification of correlated reduced
  density matrices: Review and applications},}\ in\ \href@noop {} {\emph
  {\bibinfo {booktitle} {Reduced‐Density‐Matrix Mechanics: With Application
  to Many‐Electron Atoms and Molecules}}}\ (\bibinfo  {publisher} {John Wiley
  \& Sons, Ltd},\ \bibinfo {year} {2007})\ Chap.~\bibinfo {chapter} {9}, pp.\
  \bibinfo {pages} {205--259}\BibitemShut {NoStop}%
\end{thebibliography}%
\end{document}